\documentclass[useAMS]{mn2e}

\usepackage{graphicx}
\usepackage{times}

\def\deg{$^{\circ}\,$}
\def\solm{M$_{\odot}\,$}
\def\kms{km s$^{-1}$}

\def\eg{{\it e.g.\ }}
\def\ie{{\it i.e.\ }}

\title[Signatures of orbiting MBHs in galaxies]{Can gas dynamics in centres 
of galaxies reveal orbiting massive black holes?}

\author[J. Etherington and W. Maciejewski]{James Etherington and Witold Maciejewski\\
Astrophysics, Denys Wilkinson Building, Keble Road, Oxford OX1 3RH}

\begin{document}

\maketitle

\begin{abstract}
If supermassive black holes in centres of galaxies form by merging of 
black-hole remnants of massive Population III stars, then there should be 
a few black holes of mass one or two orders of magnitude smaller than that of 
the central ones, orbiting around the centre of a typical galaxy. These black holes
constitute a weak perturbation in the gravitational potential, which can generate 
wave phenomena in gas within a disc close to the centre 
of the galaxy. Here we show that a single orbiting 
black hole generates a three-arm spiral pattern in the central gaseous disc. 
The density excess in the spiral arms in the disc reaches values of 3-12\% 
when the orbiting 
black hole is about ten times less massive than the central black hole. 
Therefore the observed 
density pattern in gas can be used as a signature in detecting the most massive 
orbiting black holes.
\end{abstract}

\begin{keywords}
hydrodynamics --- galaxies: kinematics and dynamics --- 
galaxies: spiral --- galaxies: structure --- galaxies:
nuclei --- cosmology: dark matter
\end{keywords}

\section{Introduction}
Numerical simulations of structure formation indicate that baryonic cores 
within dark matter haloes collapse at redshifts $z\approx20-30$ 
to sufficiently high densities to be available for primordial 
star formation (\eg Abel et al 1998). Because of the lack of metals 
that could facilitate cooling, the collapse occurs without 
much fragmentation and it gives rise to the Population III stars, much 
more massive than stars forming today (Abel, Bryan \& Norman 2000). 
Stars as massive as $10^3$ \solm could 
form, and they could evolve into massive black holes (MBHs) with little mass loss 
(Madau \& Rees 2001). A Population III star with a mass greater than 
260 \solm will not experience a supernova explosion, because the gravity of 
the star is too strong, and instead it will simply collapse to form a black 
hole. Islam, Taylor \& Silk (2003, 2004a) followed in detail hierarchical merging of such a 
population of primodial black holes, and they found that about $10^3$ MBHs 
would be present in the galactic halo today. In total, mass comparable or greater 
than that of the central supermassive black hole (SMBH) should still reside in 
MBHs orbiting in galactic halos. Most of 
these MBHs would have masses near the initial seed mass of the remnant Population III star, 
but there would be a few as massive as $10^6-10^7$ \solm.

In this paper we investigate whether the most massive MBHs orbiting in galactic halos can
be revealed by structures related to waves that they generate in gaseous discs 
in centres of galaxies. At the nuclear scales, gas dynamics becomes decoupled 
from dynamics of the stars, and one can study the response the of the former to any imposed 
gravitational potential. A linear theory applicable here was developed by Goldreich \& 
Tremaine (1978, 1979), who showed that density waves can propagate within the gas as a response 
to a rigidly rotating potential. The propagation of 
waves depends crucially on the resonances created by the rotating potential, since 
waves are generated at these resonances. 
Waves propagating in gas generate spiral structure there, 
and the linear theory was first invoked by Lindblad \& J\"{o}rs\"{a}ter (1981) to explain 
the spiral structure observed in the centre of a barred galaxy. Maciejewski (2004 a,b) used 
the linear theory and hydrodynamical modelling to analyse nuclear spirals generated by a wide 
range of galactic
potentials, including ones that have a SMBH at the centre 
and a bisymmetric bar.
These latter papers suggest that asymmetries in the galactic potential too weak 
to be detected observationally might be sufficient to generate nuclear spirals 
that should be easily observed. 

A single MBH orbiting in the galactic halo gives rise to
an asymmetric perturbation in the potential rather than to a bisymmetric one, caused
by a bar, that has been studied previously. In this paper we investigate in detail gas 
dynamics generated by this potential using hydrodynamical modelling. There are similarities
here to the interaction between a protoplanet and the protostellar disc (e.g. Lin \& 
Papaloizou 1993), although there are also significant differences which we point out.

The structure of the paper is as follows. Section 2 presents the numerical code, the gravitational 
potential and the parameters used. All of the models are summarized in Table 1, and in Section 3 
representative models are analyzed further. An interpretation of the results is 
given in Section 4. Implications of the models for the detection of MBHs orbiting in galaxies, and 
other detection methods are discussed in Section 5. 
We summarize our conclusions in Section 6.

\section{Methodology}
The modelling of gas dynamics was performed using the CMHOG2 code that was developed from 
the original CMHOG code by Piner, Stone \& Teuben (1995). The code solves the hydrodynamical fluid 
equations on a log-polar grid in two dimensions. The PPM scheme is employed and the 
gas is assumed to be isothermal. 
Each time-step the code calculates three hydrodynamical variables: the radial and tangential 
components of the gas velocity and  
the gas density. By modifying appropriate subroutines it is possible to model the gas in 
practically any gravitational potential desired. The data files that the code creates can be used 
to study the time evolution of the gas, investigate the structure and dynamics of gas and search for  
stable patterns in gas flow. The log-polar grid extends from an 
inner radius of 5 pc to an outer radius at 4 kpc, which is large enough to encircle 
all wave phenomena. At the outer boundary the waves have effectively 
decayed completely so reflections there are insignificant. We use closed boundary (reflection) 
conditions at both the inner and outer ends of the grid. The grid consists of 174 radial 
annuli that are split by 160 angular slices. The grid is logarithmic, with 
the smallest cells at the inner boundary, to make the cells approximately square. 
The cell size at the inner 
boundary is 0.19 pc, which allows the central region to be examined in great detail, and at 
the outer boundary it is 160 pc. 

\begin{table*}
 \caption{29 models with their parameter values and extracted characteristics:  
(1) model name, (2) number of orbiting MBHs, (3) inclination of the MBH's orbit 
with respect to the galactic plane, (4) logarithm of the mass of the MBH in \solm, 
(5) softening in Plummer's 
formula for orbiting MBHs, 
(6) central gas morphology and stability of wave pattern (for details see Section 3), (7) maximum 
arm-interarm density 
ratio (density measured in \solm pc$^{-2}$ at radii larger than 100 pc) and its corresponding 
density excess in spiral arms in brackets (global values are given if no structure) and 
(8) time at which densities were measured.}
 
 \begin{tabular}{@{}rcrcrccr}
 \hline 
    Name & Number of MBHs & i & log(M) & r$_s$ [pc] & Morphology & Max Density Ratio & Time [Myr]  \\
  (1)&(2)&(3)&(4)&(5)&(6)&(7)&(8) \\
 \hline 
   1 & 1 & 0 & 6 & 100 & disturbed 3-arm & 10.02/9.94 (0.8\%) & 100   \\                 
   2 & 1 & 0 & 7 & 10 & disturbed 3-arm & 10.58/9.65 (9.6\%) & 100   \\              
   3 & 1 & 0 & 7 & 50 & disturbed 3-arm & 10.53/9.45 (11.4\%) & 100   \\               
   4 & 1 & 0 & 7 & 100 & disturbed 3-arm & 10.58/9.79 (11.9\%) & 100 \\                 
   5 & 1 & 0 & 8 & 100 & disturbed 3-arm & 14.41/8.67 (66.2\%) & 100  \\               
   6 & 1 & 45 & 6 & 100 & no structure & 10.01/9.98 (0.3\%) & 100\\              
   7 & 1 & 45 & 7 & 10 & variable 3-arm & 10.10/9.86 (1.9\%) & 75 \\   
     &   &    &   &    &                & 10.10/9.86 (2.4\%) & 100  \\ 
   8 & 1 & 45 & 7 & 50 & variable 3-arm & 10.06/9.91 (1.5\%) & 75  \\  
     &   &    &   &    &                & 10.05/9.92 (1.3\%) & 100 \\  
   9 & 1 & 45 & 7 & 100 & variable 3-arm & 10.04/9.93 (1.1\%) & 75\\  
     &   &    &   &     &                & 10.12/9.80 (3.2\%) & 100  \\  
  10 & 1 & 90 & 6 & 100 & no structure & 10.01/9.97 (0.4\%)  & 100 \\              
  11 & 1 & 90 & 7 & 10 & variable 3-arm & 10.07/9.82 (2.5\%) & 72  \\    
     &   &    &   &    &                & 10.14/9.79 (3.5\%  & 100 \\
  12 & 1 & 90 & 7 & 50 & variable 3-arm & 10.11/9.86 (2.5\%) & 72  \\    
     &   &    &   &    &                & 10.13/9.82 (3.1\%  & 100  \\
  13 & 1 & 90 & 7 & 100 & variable 3-arm & 10.15/9.82 (3.3\%) & 72 \\    
     &   &    &   &     &                & 10.07/9.89 (1.8\%) & 100 \\
  14 & 2 & 0 & 6 & 100 & stable 4-arm & 10.04/9.95 (0.9\%) & 100 \\            
  15 & 2 & 0 & 7 & 10 & stable 4-arm & 10.53/9.50 (10.8\%) & 100 \\            
  16 & 2 & 0 & 7 & 50 & stable 4-arm & 10.66/9.55 (11.0\%) & 100 \\            
  17 & 2 & 0 & 7 & 100 & stable 4-arm & 10.61/9.53 (11.3\%) & 100 \\           
  18 & 2 & 0 & 8 & 100 & stable 4-arm & 11.60/8.95 (29.6\%) & 100 \\           
  19 & 2 & 45 & 6 & 100 & no structure & 10.00/9.97 (0.3\%) & 100 \\       
  20 & 2 & 45 & 7 & 10 & stable 2-arm & 10.26/9.64 (6.4\%) & 100 \\           
  21 & 2 & 45 & 7 & 50 & stable 2-arm & 10.18/9.70 (4.9\%) & 100 \\           
  22 & 2 & 45 & 7 & 100 & stable 2-arm & 10.12/9.81 (3.1\%) & 100 \\          
  23 & 2 & 45 & 8 & 100 & stable 2-arm & 11.40/9.12 (25.0\%) & 100 \\         
  24 & 2 & 75 & 7 & 100 & variable 4-arm & 10.07/9.91 (1.6\%) & 100 \\              
  25 & 2 & 90 & 6 & 100 & no structure & 10.00/9.97 (0.3\%) & 100 \\            
  26 & 2 & 90 & 7 & 10 & variable 4-arm & 10.00/9.93 (1.4\%) & 58 \\   
     &   &    &   &    &                & 10.14/9.82 (3.2\%) & 89 \\        
  27 & 2 & 90 & 7 & 50 & variable 4-arm & 10.06/9.92 (1.4\%) & 58 \\   
     &   &    &   &    &                & 10.12/9.80 (3.2\%) & 89 \\        
  28 & 2 & 90 & 7 & 100 & variable 4-arm & 10.08/9.92 (1.61\%) & 58 \\         
     &   &    &   &     &               & 10.12/9.85 (2.7\%) & 89 \\       
  29 & 2 & 90 & 8 & 100 & variable 4-arm & 10.72/9.51 (12.7\%) & 58 \\                
     &   &    &   &     &               & 11.46/8.66 (32.3\%) & 89 \\       
   
 \hline  \\
\end{tabular}
\end{table*}

The code uses Athansasoula`s (1992) model of the galactic potential which 
originally had two components: the disc and the bulge. Following Maciejewski (2004b) we 
added a central SMBH. The disc has a surface density of the form: 
$\sigma(R)=\sigma_0(1+R^2/R_0^2)^{-3/2}$, which requires two parameters: the surface density 
at the centre of the disc $\sigma_0$ and a
characteristic radius $R_0$. The bulge has a volume density: $\rho(r)=\rho_b(1 + r^2/r_0^2)^{-3/2}$. 
Similarly, it requires two free parameters: the density at the centre of the bulge $\rho_b$ 
and a characteristic
radius $r_0$. All four parameters are taken from Athansasoula's standard model. 
The SMBH is modelled using Plummer`s formula, which ensures that there is no 
singularity at its location. 
The mass of the SMBH is 10$^8$ \solm throughout this work, which corresponds to 0.65\% 
of the mass within the inner 1 kpc from the centre, 
and the softening is 0.01 kpc. 

For the purpose of this work a subroutine was created 
to simulate either one or two 
MBHs orbiting the galactic centre. 
As a first 
step, we studied gas response to two equal MBHs, one being a mirror symmetry of the other. 
This situation, although unphysical, can be directly compared to simulations of gas flow in bars
because of its bisymmetry. In all our models one or two MBHs follow a circular orbit with a 1 kpc
radius. The MBH mass was 
set to 10$^6$, 10$^7$ or 10$^8$ \solm. The first two values correspond to the most massive MBHs 
that may orbit around the centres of galaxies. The last value, although unphysical, was used to 
estimate how the density contrast in gas changes with the MBH mass. Plummer`s formula was 
used again to soften the MBH mass distribution, but this time the softening was set set to 
10, 50 or 100 pc. On the grid used, the cell size at 1 kpc is 40 pc. The orbiting MBHs are 
not permitted to accrete gas in the sense that there is no loss of gas from the grid at the 
location of the MBHs. However, in the softened potential of a Plummer sphere, there is still
a possibility that the disc gas may be captured by the orbiting MBH.

The hydrodynamical simulations were initially performed in a frame 
rotating with the MBHs located in the galactic plane.
Later the code was modified to follow the MBHs orbiting in a stationary frame. The second method was 
advantageous since it allowed us to study orbital motion out of the plane of the galaxy, for which 
the projection of the MBH on the galactic plane moves with changing angular speed. To check for 
consistency, models with MBHs 
orbiting in the galactic plane were created with the two versions of the code.
The outputs showed no detectable differences. 

The orbital period at a radial distance of 1 kpc is about 20 Myrs. To ensure that the 
response of the gas was smooth, the mass of the off-centre MBHs was 
introduced gradually during the first orbit. To obtain the steady state response of the 
gas the models were run for 100 Myrs. This 
corresponds to about 5 orbits, which was long enough to recover all relevant structures in gas. 
The sound speed in gas in the models is 20 \kms, which is typical for gas velocity dispersion in the inner 
parts of galaxies. A wave in such gas can propagate a distance of 2 kpc in 100 Myrs. 
Thus the wave has enough time to 
cross the area encircled by the MBH's orbit within the time of the run.       

\begin{figure*}
\centering
\vspace{-25mm}
\includegraphics[width=\linewidth]{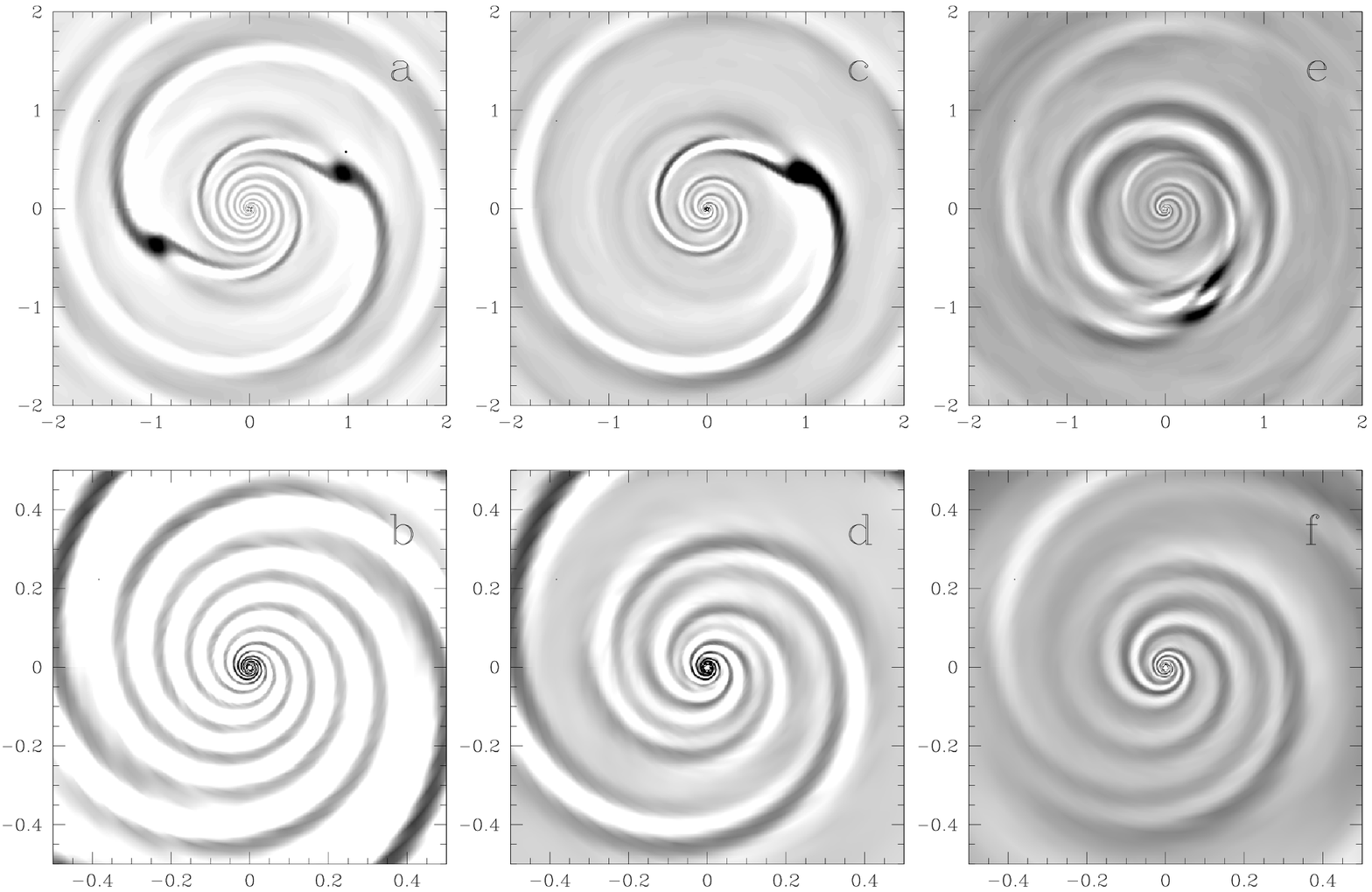}
\vspace{-105mm}
\caption{a) The density of gas in the galactic plane in greyscale for model 17, 
which has two 10$^7$ \solm MBHs with softening of 100 pc orbiting in the plane of the galaxy, seen 
at 100 Myrs. 
Darker shading represents areas of larger density. c) Same as for a) but for model 4, which 
has one 10$^7$ \solm MBH with softening of 100 pc orbiting in the plane of the galaxy, seen at 100 Myrs.
e) same as a) but for model 11 which has one 10$^7$ \solm MBH with softening of 10 pc 
orbiting perpendicular to the 
plane of the galaxy, seen at 70 Myrs. Panels b), d) and f) show details 
for the central kpc of corresponding panels a), c) and e). Units on the axes are in kpc.} 
\label{f1}
\end{figure*}

\section{Results}
In total we built 29 models, for which we varied four parameters: the number of orbiting MBHs, 
the inclination of their orbit with respect to the galactic plane, as well as the mass and softening 
of the orbiting MBHs. The models together with their parameters are presented in Table 1. Below we 
analyze representative models. 

\subsection{Models with two orbiting MBHs in the galactic plane}
Fig. 1a shows the gas density in the galactic plane for model 17, in which two 10$^7$ \solm MBHs are
orbiting in the plane of the galaxy. The wave pattern stabilizes at about 60 Myrs and there are no 
significant changes after this time. In Table 1 this pattern is referred to as 'stable 4-arm'. 
The two very dark regions indicate high densities of gas 
captured by the two orbiting MBHs. At this radius (1 kpc) the two MBHs drive a strong two-armed spiral. 
Inside the radius of about 600 pc the two-armed spiral 
changes into a spectacular four-armed spiral. The maximum density excess in the arm of the 
four-armed spiral is about 11\% at a radius of 260 pc.  
There is a larger density excess at radii smaller than 100 pc, but 
the values presented here and in Table 1 are measured for radii larger than 100 pc, since spiral structure 
of smaller extent is unlikely to be resolved in observed galaxies.  
Fig. 1b shows that the four-armed wave pattern continues to the innermost parsecs of the galaxy 
and that it winds tightly around the centre. The formation of the spiral indicates that
the waves propagate towards the centre of the disc, and in our model they are reflected 
from the inner boundary of the grid. The reflected waves interfere with the incoming
ones, but because they geometrically diverge, they are quickly weakened as they move 
away from the boundary.

Radial and tangential density profiles for model 17 are shown in Fig 2. The radial 
plot indicates that between the radii of 50 pc and 0.4 kpc the wave pattern is rather uniform with a 
maximum density of about 10.5 \solm pc$^{-2}$ and a minimum density of about 9.5 \solm pc$^{-2}$. 
Inflow makes the density larger at smaller radii.
The plot of the tangential density profile shows that the multiplicity of the wave pattern is 
clearly four and the amplitude of all four arms is about the same. 

The density contrasts and 
the morphologies for other models are given in Table 1. It is useful to compare model 17 to 
models 14 and 18, where the masses of the orbiting MBHs
are 10$^6$ \solm and 10$^{8}$ \solm respectively. The 10$^6$ \solm MBHs are too small 
a perturbation to generate density waves with a significant amplitude, and the density excess in the spiral 
arms for the 10$^{8}$ \solm MBH model is just below 30\%. The values of density excess in all three 
models indicate that it 
appears proportional to the MBH mass in the range between 10$^6$ and 10$^7$ \solm 
(and perhaps below it), but 
it increases more slowly for masses above 10$^7$ \solm, 
most likely because of reaching the non-linear regime.

Parameters of models 15--17 are identical except for the softening in Plummer's formula for the MBH. 
The density contrast appears to increase slightly 
as the softening value decreases making the images sharper. However the effect is small 
and we do not analyze it any further.

\begin{figure}
\centering
\vspace{-5mm}
\includegraphics[width=\linewidth]{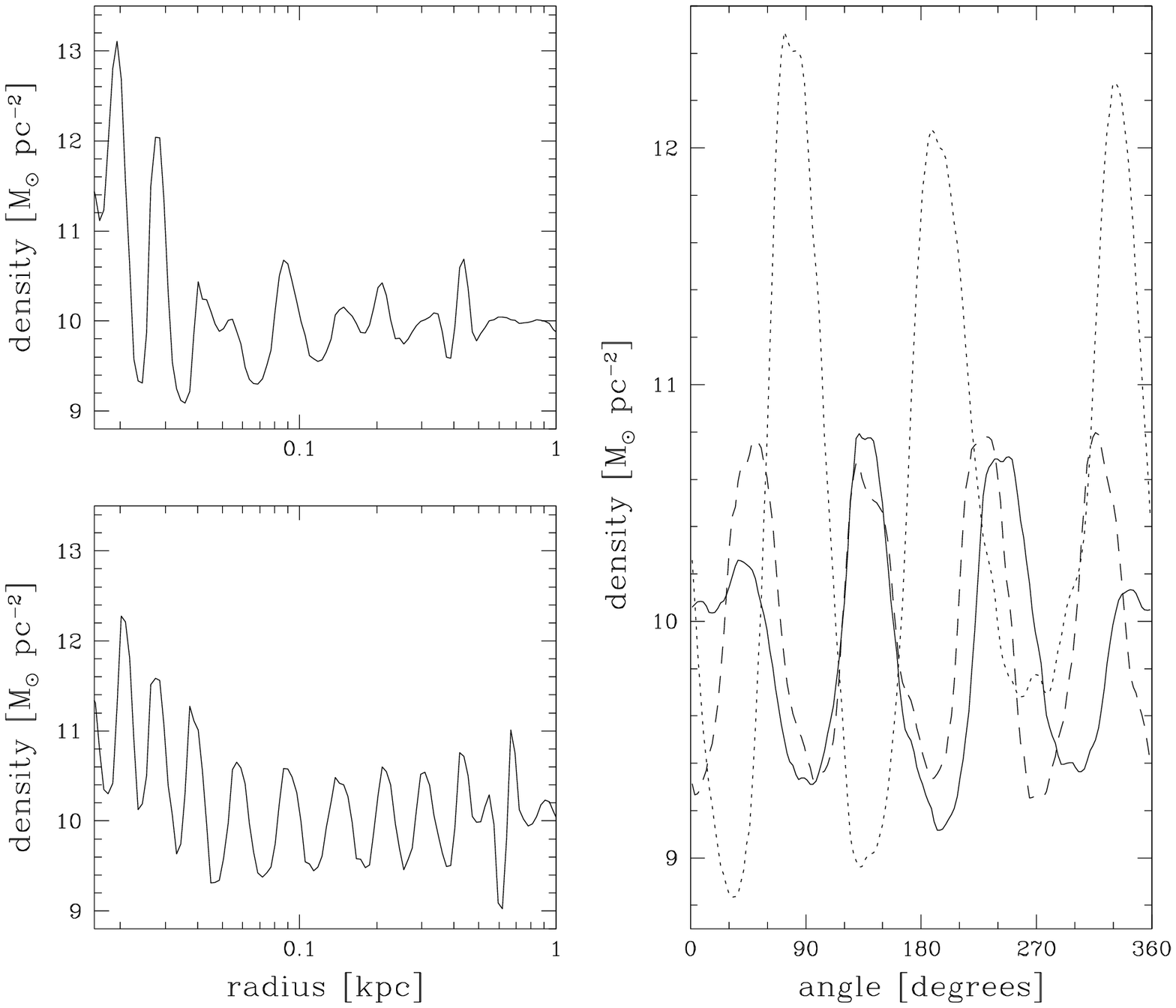}
\vspace{-11mm}
\caption{{\bf Left:} Gas density profiles as a function of radius, taken at 100 Myrs at an angle 
of 80\deg clockwise with respect 
to the positive $x$-axis. The top panel represents model 4 with one orbiting MBH and the bottom 
panel model 17 with two orbiting MBHs. 
{\bf Right:} Gas density profiles as a function of angle taken at 100 Myrs. 
The solid and dotted curves mark the profiles for 
model 4 taken at a radius of 71 pc and 29 pc, respectively. The dashed curve is 
the profile for model 17 taken at a radius of 71 pc.}
\label{f2}
\end{figure}

\subsection{Models with two MBHs in orbits inclined to the galactic plane}
Each of the models 20, 24 and 26 has two 10$^7$ \solm MBHs orbiting in a plane 
inclined to the galactic plane. The inclination angle is 45\deg, 75\deg and 
90\deg respectively. A stable two-arm spiral within a set of rings is generated in model 20. The maximum 
density excess in the spiral arms is 6.4\% at a radius of 170 pc. The rings outside the spiral arms 
are disturbed periodically by the motion of the orbiting MBHs. In model 24 a four-arm spiral is seen. 
It has a maximum density excess of only 1.6\% and the density contrast changes 
considerably along the spiral arms, with the spiral morphology evolving in time. In model 26 
the waves generated in gas 
produce structures that change in time. At 
60 Myrs a uniform four-arm spiral is seen with a density contrast 
of 1.4\%. This subsequently evolves 
into a four-armed structure that has two strong arms and two weaker arms. 
Its density contrast at about 90 Myrs is 3.2\%. In Table 1 we refer to this structure 
as 'variable 4-arm'.

\subsection{Models with one orbiting MBH in the galactic plane}
In simulations with one MBH orbiting in the galactic plane, the gas morphology is vastly 
changed. Fig. 1c shows gas density in model 4. The single very dark region 
represents a high density of gas 
captured by the MBH orbiting in the galactic plane. This MBH drives a 
strong single-armed spiral at neighbouring radii. However, at radii of less than 
450 pc a more complex behaviour is seen. There are three spiral arms emerging from the galactic 
centre with 
a maximum density excess in the arms of 12\%. However, unlike in model 17 with two MBHs, the 
density contrast here varies along the spiral arms (Fig. 2, top-left panel) and at some radii a 
fourth arm appears, like at radius about 200 pc at negative $y$ coordinates. The azimuthal 
density profile at most radii shows three maxima, one of which is broad, with two peaks (Fig. 2, right 
panel). This may be interpreted as a four-arm spiral, but contrary to the profile of
a clear four-armed spiral in model 17, the spacing of the peaks 
is uneven here. Thus the wave pattern is closer to a disturbed three-armed spiral and it is 
classified as such in Table 1. We note however, that inside the radius of 
50 pc the pattern consistently shows three clear spiral arms of equal amplitude, 
as indicated by the dotted line in the 
right panel of Fig. 2. Fig. 1d shows that the spiral pattern propagates to 
the centre of the galaxy, and that it tightly winds around it.

\subsection{Models with one MBH in orbit inclined to the galactic plane}
Model 11 (Fig. 1 e,f) has a single MBH in an orbit perpendicular to the galactic plane. The 
basic difference between this model and model 4, with a MBH in the galactic plane is that 
the strong single-arm spiral present in the latter is missing now. 
Now, the orbiting MBH that passes through the 
galactic plane, causes a disturbance to what otherwise looks like a set of rings
of radii similar to that of the MBH's orbit. The wave pattern now becomes more time 
dependent and the disturbance in the rings is reinforced each time the black hole passes through the plane. 
However, the wave pattern generated in the central region of the galaxy, 
within a radius of about 300 pc, does not 
show periodic changes related to the orbiting MBH, and for most 
of the run it is importantly still a 
three-armed spiral. The maximum density excess in the spiral arms is 
about 3\% at radii above 100 pc, but it 
gets larger further in, reaching 5\% and more. 
Towards the end of the run, at 100 Myrs, this pattern evolves 
into a two-arm spiral with a weak third arm. We refer to this as a 
'variable 3-arm' spiral in Table 1. 

Model 7 has a single MBH orbiting in a plane inclined at 45\deg to the galactic plane. 
The morphology in this model is the same as seen in model 11, but 
the density contrast in the spiral is weaker than in model 11, while it is stronger in the rings 
outside the spiral.    
 
\section{Interpretation}
We expect that the spiral structure in gas in the centres of galaxies, 
observed in our models, is the consequence of waves generated 
by orbiting MBHs. In the linear theory of density waves in gas resonance-excited 
by a rotating asymmetry in the potential, in the absence of self-gravity
(Goldreich \& Tremaine 1978, 1979), an $m-$arm 
spiral is generated at the resonance, where the epicyclic 
frequency $\kappa$ is integer $m$ times larger than the orbital frequency 
$\Omega$ in the frame rotating with the asymmetry. Thus the condition for a 
resonance inside the orbit of the MBH is $\Omega - \kappa/m = \Omega_p$, where $\Omega_p$ is 
pattern speed of the asymmetry --- in our case the angular velocity of the MBH. Spirals 
generated by such resonances are confined within their radii.

In the first part of this work models with two orbiting 
MBHs were constructed. Although such a setup is unphysical, the forcing 
here is bisymmetric, like in the well established models of gas flow in
barred galaxies. By comparing our results to those models we can make
first steps in the interpretation of the morphology that we observe.
Because of bisymmetry, gravitational potentials in both cases do not contain odd modes,
hence they cannot trigger odd-multiplicity spirals, which is confirmed by 
the models. However, a four-arm spiral is seen in the central parts of the 
galaxy in our models, contrary to a two-arm spiral which appears in models with 
bars. Why does a pair of orbiting MBHs generate a four-armed spiral while a two-armed spiral is 
observed in the case of a bar?    
To explain this difference, we made plots of $\Omega - \kappa/m$ 
against radius for various $m$ for the gravitational potential used in our models. 
In the case of two MBHs orbiting in the 
galactic plane the $m=4$ resonance is located at 620 pc. This agrees 
remarkably well with the distance from the centre, at which 
the four-arm spiral ends in Fig. 1a. In the linear theory a four-arm spiral should 
be generated by a $m=4$ resonance and propagate inwards, and the models indicate that 
this mechanism indeed operates in galactic centres: the spiral density waves are being driven 
by the MBHs that constitute an asymmetric perturbation in the galactic potential. 

\begin{figure}
\centering
\includegraphics[width=\linewidth]{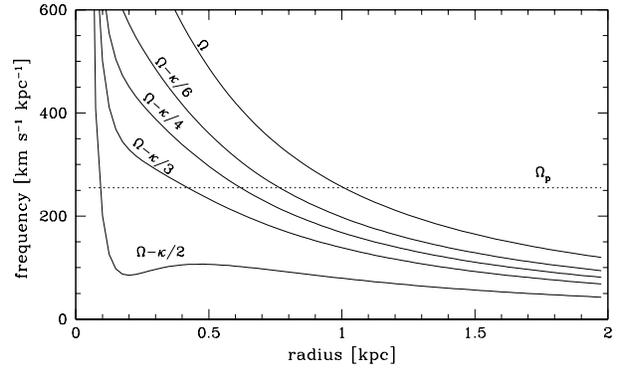}
\vspace{-35mm}
\caption{Frequency-radius diagram for the gravitational potential used in our models. 
$\Omega - \kappa/m$ curves are represented by solid lines. The dotted line marks the pattern speed 
$\Omega_p$ corresponding to the angular velocity of a MBH placed at a radius of 1 kpc.}
\label{f3}
\end{figure}

However, in the gravitational potential of our models 
an $m=2$ resonance is present at the radius of 93 pc (Fig. 3). Why is a two-armed 
spiral not generated there? Here understanding can be gained by comparing the ratio 
of the tangential to radial components of the gravity force between our model with orbiting 
MBHs, and a potential of a weak oval, which is 
known to generate two-arm spirals (Maciejewski 2004b). Contour plots for these ratios are presented in Fig. 4. 
Note that in the case of the oval, the maximum tangential 
forcing is well inside the corotation radius. On the other hand, in the case of an 
orbiting MBH the tangential forcing is largest at corotation, and it 
quickly loses its strength at smaller radii, where resonances are located. 
In the case of a weak oval 
(Fig. 4, right panel) a force ratio of 0.006 at the $m=2$ resonance is sufficient 
to drive the two-arm spiral mode. This mode prevails despite 
the fact that at the position of the $m=4$ resonance the force ratio is 
about 50\% larger at 0.01. On the other hand, in the case of two orbiting 10$^8$ \solm MBHs 
(Fig. 4, left panel) the force ratio at the $m=2$ 
resonance is 0.001 whereas at the $m=4$ resonance it is 0.014. Reduction of tangential forcing at the 
$m=2$ resonance by a factor of 14 is likely the reason why the two-arm mode is not observed in our models. 
Relatively weak tangential forcing at low multiplicity modes is an inevitable feature of 
potentials with orbiting 
bodies, as Fig. 4 indicates.

Our interpretation of structures observed in models with two orbiting MBHs was aided by the 
existing work on bisymmetric potentials. After learning from this case, we move to 
the physically relevant situation, where there is one orbiting MBH, corresponding to the most
massive MBH in the distribution predicted by Islam et al. (2003, 2004a). In this case, 
the gravitational potential 
is no longer bisymmetric, so the presence of a spiral with an odd number of arms is not unexpected. 
However, in the linear theory the $m=1$ mode cannot be driven, because $\Omega - \kappa$ is always 
negative, hence there is no $m=1$ resonance. The single arm seen in our models 
at radii close to the MBH (Fig. 1c) is 
a nonlinear effect, or an interference of higher-order modes (Ogilvie \& Lubow 2002).
The next odd mode, $m=3$, generates a three-armed spiral. The $m=3$ resonance is found 
to be at the radius of 422 pc, and this agrees very well
with the extent of the three-arm spiral observed in the representative model 4. However, the three-arm 
spiral seen in this model is not completely 
uniform, and it exhibits some even-multiplicity structure. From Fig. 4 we see that the maximum tangential 
to radial force ratio at the $m=3$ resonance is 0.005, while 
at $m=4$ resonance it is 0.014, less than three 
times larger. Thus one should expect contributions to the structure in our models 
from even-multiplicity modes. This analysis shows that the linear theory is applicable 
also to the case of a single orbiting MBH and that this MBH is the driving perturbation in this case.

\begin{figure}
\centering
\vspace{3mm}
\includegraphics[width=0.48\linewidth]{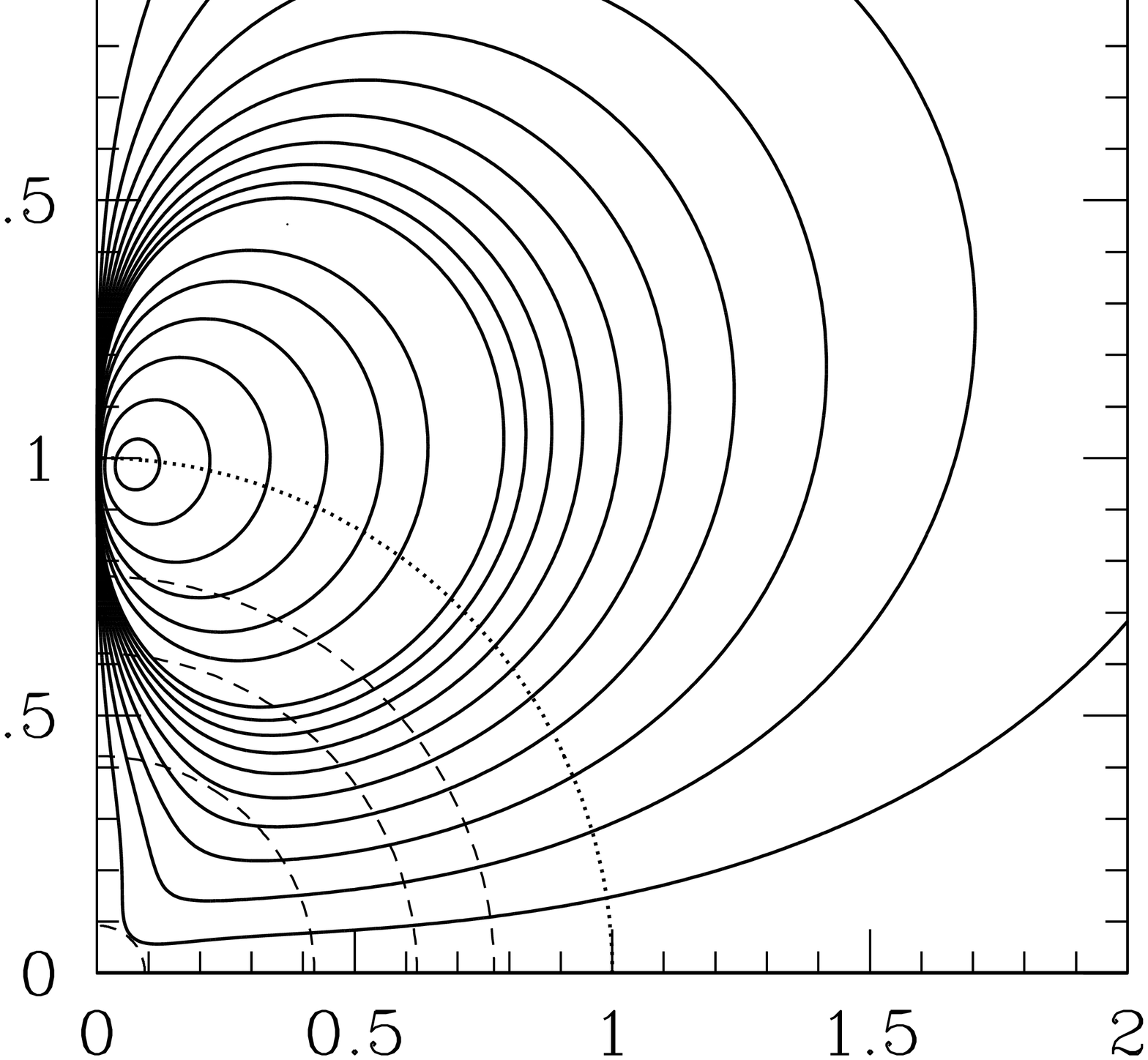}
\includegraphics[width=0.48\linewidth]{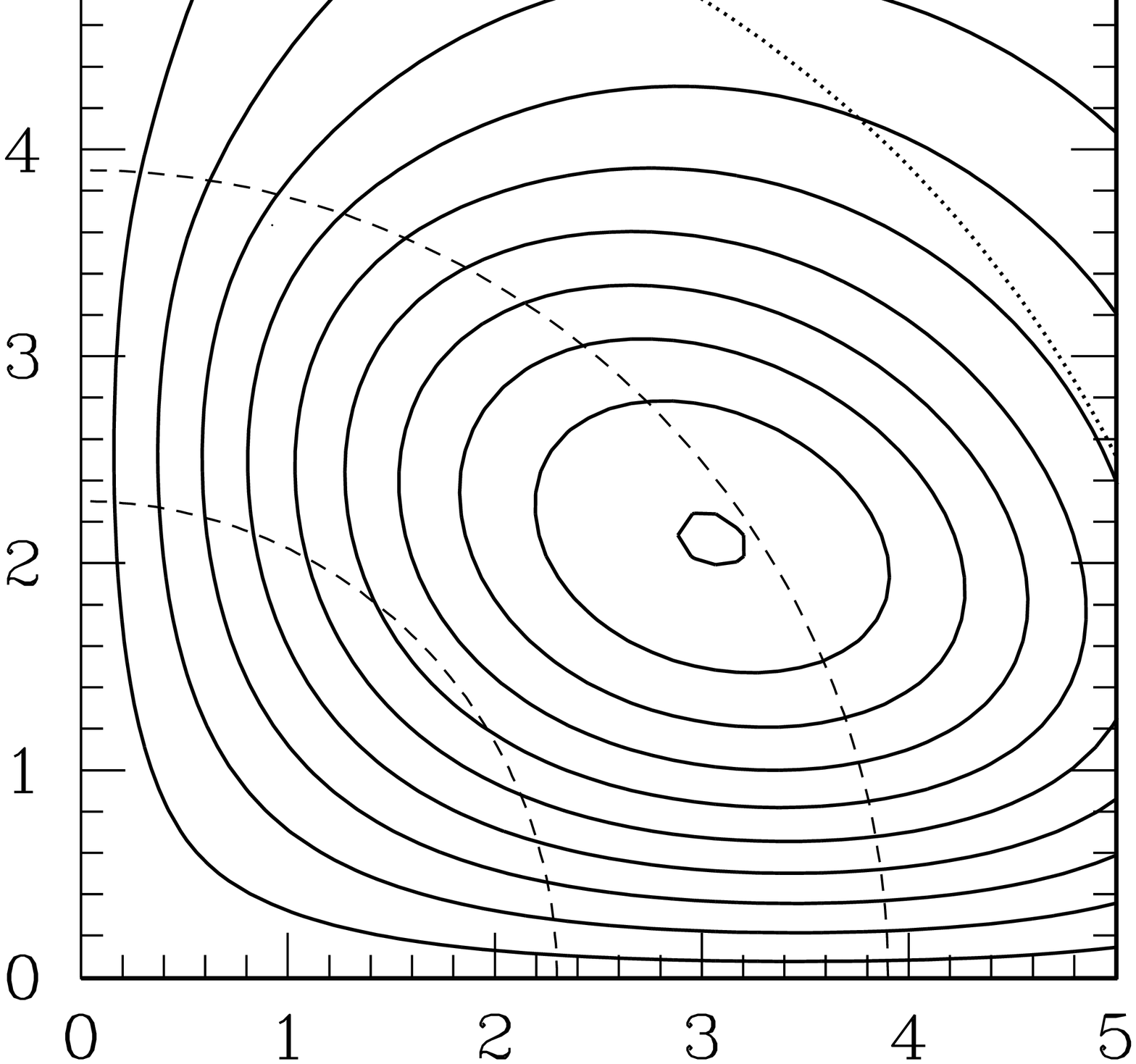}
\vspace{-12mm}
\caption{The solid lines mark contours of constant ratio of the tangential to 
radial components of the gravity force. The case of our potential 
with a 10$^8$ \solm MBH softened at 100 pc, placed on the 
$y$-axis at 1 kpc is presented in the left panel. An oval distortion 
used by Maciejewski (2004 b) is shown in the right panel. 
The contour values, from outer to inner, are: 
0.001 to 0.01 at 0.001 intervals, then 0.015, 0.02, 0.03, 0.05, 
0.1 and 0.2. The dashed curves represent the positions of the resonances. 
On the left panel, the $m=2$ resonance is at 93 pc, $m=3$ at 422 pc, $m=4$ at 621 pc and $m=6$ 
at 771 pc. The dotted line marks the corotation at 1 kpc. On the right panel, the $m=2$ resonance 
is at 2.3 kpc, the $m=4$ 
at 3.9 kpc and the corotation at 5.9 kpc is marked by the dotted line. 
The units on the axes are in kpc.}
\label{f4}
\end{figure}

\section{Discussion}
\subsection{Upper mass limit and the number of orbiting MBHs}
The work reported here investigated the response of gas in the central parts 
of a galactic disc to the periodic forcing from possibly most massive orbiting
MBHs. Evolutionary merging scenarios predict that MBHs as massive as $10^7$
\solm can form in galactic halos (Islam et al 2003, 2004a). 
Many of these most massive MBHs will spiral into the nucleus within the Hubble 
time because of dynamical friction. From formula (7-26) in Binney \& Tremaine 
(1987) we get that a $10^6$ \solm MBH, which formed at a distance of
1 kpc from the galactic centre, will spiral in within less than 10 Gyr. Since
the friction timescale is roughly inversely proportional to the orbiting mass,
the upper mass limit for a MBH formed at 1 kpc and still residing there is 
below $10^6$ \solm. However, MBHs were likely to form at larger radii. Since 
the friction timescale is also proportional to the square of the initial 
radius, a $10^7$ 
\solm MBH which formed at a 5 kpc radius will need about 20 Gyr to spiral in,
out of which it will take almost 1 Gyr to move through the inner kpc. This last value
is still much larger than the orbital period at 1 kpc, which is about 20 Myr.
Thus a $10^7$ \solm MBH spends enough time at the radius of 1 kpc to generate
the structure in gas analyzed in this work. On the other hand, a $10^8$ \solm 
MBH will need only less than 100 Myr to spiral in from 1 kpc to the galactic
centre, so there will be no steady state gas response. Anyway, such accretion 
events will be too rare to be observed.

There are other constraints on the upper mass of the orbiting MBHs. MBHs that
sink to the galactic centre through dynamical friction most likely contribute 
to the formation of a SMBH there. Unacceptable build-up of this central object 
would occur if the halo were to consist of MBHs with all masses higher than 
$\sim 3 \times 10^6$ \solm (Xu \& Ostriker 1994). Also stellar velocity 
dispersion in galactic discs would increase beyond the observed 
values if bodies orbiting in the halo were too massive. For a galaxy like the Milky Way 
it would happen when all MBHs were to have mass larger than 
$\sim 3 \times 10^6$ \solm (Carr \& Sakellariadou 1999). This upper mass limit 
can be pushed up for more massive galaxies. From all the constraints above, 
the upper mass limit for orbiting MBHs should be somewhere in the range of a few 
times $10^6$ \solm (see also van der Marel 2004 for a review).

An important limitation of Islam et al. (2004a) simulations, from which we got
the predicted population of orbiting MBHs, was that their 
orbital calculations could not be followed down to arbitrarily small radii. 
If the orbiting MBH reached a distance of 3 kpc from the centre, it was 
assumed that it had `fallen in`. Thus all MBHs that were less than 3 kpc away 
from the centre were assumed to merge efficiently with one another and with 
the central SMBH. This is quite a severe limitation considering that the 
Schwarzschild radius for a 10$^8$ \solm SMBH is only about 400 R$_{\odot}$. 
Simulations by Milosavljevi\'c \& Merritt (2001) indicate that the
separation between two MBHs at the centres of merging galaxies remains 
approximately constant at around 0.3 kpc for Milky Way sized galaxies
for a significant time before the MBHs form a hard binary. It is therefore 
conceivable that MBHs could orbit around the centres of galaxies at radii 
much smaller than 3 kpc, and the number of orbiting MBHs should therefore 
be larger than what is indicated by Islam et al. (2004a).

\subsection{Detectability of orbiting MBHs through structures in gas}
Phenomena studied in this work are similar to tidal interactions in protoplanetary
systems (e.g. Lin \& Papaloizou 1993, Bryden et al. 1999), although there
are significant differences between the two. Discs considered here are not 
Keplerian, and their interstellar medium is best described by warm gas (e.g.
Englmaier \& Gerhard 1997). Orbiting MBHs considered in this work spiral into
the galactic centre within few tens of their orbital period, and therefore 
features observed in our models are relevant only if they occur on similar
timescale. For this reason we do not expect in galactic discs a gap-opening,
similar to that occurring in protoplanetary discs after a few hundreds of
protoplanet's orbits. Moreover, orbiting MBHs are not confined to the galactic
plane, and off-plane MBHs are much less efficient in gap-opening in the disc.
The common feature of protoplanetary discs and galactic discs with MBHs in
the disc plane, the one-arm spiral that accompanies the orbiting body, is also
absent in the case of MBHs not confined to the galactic plane. However, the
spiral structure created in gas in the central parts of the galactic discs,
at radii at least twice smaller than the orbital radius of the MBH, is of 
similar strength for MBHs orbiting in the disc plane and out of the plane. 
This is the structure on which we focus in this paper.

MBHs not confined to the galactic plane will generate vertical modes in the gaseous
disc. We expect these modes to be of similar magnitude to the horizontal modes,
which are weak and remain within the linear regime. Thus the full solution is
likely to be a linear superposition of the modes, out of which we consider here
only those propagating in the disc plane. These modes are most important when
interpreting the observed nuclear morphology in galaxies. However, full
three-dimensional simulations, beyond the scope of this paper, are required 
to explore the possible coupling of horizontal and vertical modes.

Will orbiting MBHs of the highest permittable mass generate spiral structure in gas 
that is observable with currently available techniques? Dusty filaments that 
have been discovered recently in the centres of galaxies (e.g. Martini et al. 
2003, Prieto, Maciejewski \& Reunanen 2005) have luminosity lower from their
surroundings by $5-10$\%. Our models indicate that density excess in the spiral
pattern generated by a $10^7$ \solm MBH ranges from
12\% for the MBH orbiting in the galactic plane to 3\% for the off-plane MBH. Thus
structures in gas generated by MBHs of this mass should be detectable, 
especially for MBHs orbiting close to the galactic plane. On the other
hand, the density excess in models with an orbiting $10^6$ \solm MBH 
is only about 1\%. A pattern of this strength will most likely remain 
undetected, and it probably will be overridden by random density 
fluctuations in centres of galaxies.

In the models constructed in this work, we assumed that the MBHs are orbiting
in the otherwise rigid potential. One should note though that the MBH masses 
considered here are large enough to generate a significant response in the 
distribution of stars. Appropriate implementation of this effect requires 
full N-body simulations, which are beyond the scope of this paper, but it is 
clear that the stellar response will amplify waves driven in 
gas by the orbiting MBH. This should enable detections of MBHs with slightly
lower masses, down to a few times $10^6$ \solm. On the other hand, the mass of
the central SMBH constitutes only 0.65\% of the total mass encircled by the 
MBH's orbit, hence no significant reflex motion of the central SMBH is expected.

Considerations above lead to an interesting conclusion, namely that orbiting MBHs 
of the highest permittable mass generate phenomena in gas that are right on 
the verge of detection with current 
observational techniques. This is not only encouraging, but it also ensures that 
in a given galaxy only one exceptionally massive MBH (if any at all) will 
generate structure in gas, and this structure therefore should be coherent. If there 
was strong response in gas to orbiting MBHs of masses orders of magnitude 
smaller than the upper mass limit, then the structure in gas would be incoherent, 
especially that gas response becomes nonlinear just one order of magnitude above 
the pattern detection limit (Section 3.1).

One of our main conclusions is that one orbiting MBH generates a three-arm spiral in 
gas in the central regions of the galactic disc. Although one observation
in the near-IR does not constitute a sample of any sort, it is suspicious
that the dusty filaments in the centre of NGC 1097 form a three-arm spiral inside
300 pc radius, even if bisymmetry prevails at larger radii (Prieto et al. 2005). 
This three-arm spiral finds no satisfactory explanation,
and further observations of centres of galaxies in the near-IR are needed
to establish how common it is.

\subsection{Other methods to detect orbiting MBHs}
If orbiting MBHs accrete gas, they can be sources of X-ray emission. Any 
compact object with X-ray luminosity higher than the Eddington limit for 
isotropic accretion onto a stellar-mass black hole 
($L_{x} \geq 10^{38}$erg s$^{-1}$) is known as an ultraluminous X-ray source 
(ULX). It has been suggested (Kaaret et al. 2001) that a population of MBHs 
could account for some fraction of the ULXs. However, emission from accreted 
interstellar medium (Bondi-Hoyle accretion) is insufficient to account for 
the observed luminosity (e.g. King et al. 2001), and gas has to be present
on site, either in a binary system or in baryonic cores. Both scenarios have
problems though, since there is no known evolutionary path that produces a 
binary of the characteristics required by the first scenario, while MBHs 
which form by merging of remnants of Population III stars may be 
stripped by now of their baryonic cores in the second scenario.

Compact objects, like orbiting MBHs, can be detected through microlensing. The 
lensing duration scales as a square root of the lens mass, with a microlensing
event caused by a lens of solar mass lasting about 130 days. Therefore even 
with long-duration events, only the low end of the mass spectrum of orbiting 
MBHs can be explored (van der Marel 2004).

Formation of MBHs at high redshift can be probed in the future with 
gravitational waves, which are released by the initial collapse of 
Population III stars into MBHs, and by subsequent mergers of MBHs.
There could be as many as $10^4-10^5$ gravitational wave events per year 
that may fall within the sensitivity limits of the proposed Laser 
Interferometer Space Antenna (LISA) gravitational wave observatory (Islam, 
Taylor \& Silk 2004b). Observations of gravitational wave events could be used 
to constrain the merger history of the MBHs and to provide improved limits on 
their abundances. The collapse of the first stars into MBHs may also produce 
gamma-ray bursts. Observing them will allow the epoch of the first star 
formation to be directly detected at redshifts larger than twenty.

\section{Conclusions}
This paper was based on the innovative idea that orbiting MBHs 
could produce a signature pattern in the gas that may 
be used to identify them. The main result of this study is that a three-armed 
spiral is generated in gas when a 
single MBH orbits around the galactic centre. This structure is the consequence of 
resonance-excited waves in gas. Only MBHs at the very
end of the upper range of masses permitted by dynamical constraints and evolutionary 
scenarios, \ie with masses nearing 10$^7$ \solm, are capable to generate structure 
in gas with density contrast high enough to be observed with current techniques. 
However, given that there are not many techniques that could assure detection of high-mass 
orbiting MBHs, the signature proposed here might be well worth pursuing. 

\section*{Acknowledgements}
We would like to thank Prof. Joe Silk for bringing to our attention the
idea that very massive black holes may be orbiting in galaxies. We thank
the anonymous referee for useful comments on the similarity of our work
to that on protoplanetary systems. This work was partially supported by
the Polish Committee for Scientific Research as a research project in the
years 2004--2006.

\end{document}